\documentclass{ws-procs975x65}
\def\adeg{^{\circ}}
\def\adegp{{\rlap.}^{\circ}}
\def\amin{^\prime}

\begin{document}

\title{{\it Fermi}-LAT RESULTS ON GALACTIC PLANE $\gamma$-RAY TRANSIENT SOURCES}

\author{SYLVAIN CHATY on behalf of the {\it Fermi}-LAT collaboration}

\address{Laboratoire AIM (UMR 7158 CEA/DSM-CNRS-Universit\'e Paris Diderot) \\
Irfu/Service d'Astrophysique, Centre de Saclay,  B\^at. 709 \\
FR-91191 Gif-sur-Yvette Cedex, France\\
E-mail: chaty@cea.fr}

\begin{abstract}
The Large Area Telescope on the {\it Fermi} $\gamma$-ray Space Telescope provides unprecedented sensitivity for all-sky monitoring of $\gamma$-ray activity. It has detected a few Galactic sources, including 2 $\gamma$-ray binaries and a microquasar. In addition, it is an adequate telescope to detect other transient sources. The observatory scans the entire sky every three hours and allows a general search for flaring activity on daily timescales. This search is conducted automatically as part of the ground processing of the data and allows a fast response to transient events, typically less than a day. Most of the outbursts detected are spatially associated with known blazars, but in several cases during the first years of observations, $\gamma$-ray flares occurring near the Galactic plane did not reveal any initially compelling counterparts. This prompted follow-up observations in X-ray, optical, and radio to attempt to identify the origin of the emission and probe the possible existence of a class of transient $\gamma$-ray sources in the Galaxy. Here we report on these LAT events and the results of the multiwavelength counterpart searches.

\end{abstract}


\bodymatter

\section{Introduction}\label{introduction}
There has been one firmly established class of variable
sources in the high-energy $\gamma$-ray sky. The
Energetic $\Gamma$-Ray Experiment Telescope ({\it EGRET})
on the Compton $\Gamma$-Ray Observatory discovered
a population of variable $\gamma$-ray blazars above 100
MeV\cite{hartman:1999}. However, {\it EGRET} also left the legacy of a
large fraction of unidentified sources in the 3EG catalog.
Many of these were found at low Galactic latitudes and
believed to be Galactic in nature. {\it EGRET} established
$\gamma$-ray pulsars as a Galactic population and these
were thought to contribute to the unidentified sources.
Several studies found indications of variability in some
of the sources along the Galactic Plane (GP)\cite{torres:2001,nolan:2003}, a
behavior not expected of the pulsars, which are steady
on these timescales. Additionally, no blazar counterpart
was identified in several cases. This suggested the
possible existence of a new Galactic $\gamma$-ray class.
Several sources showed both strong variability and a
convincing lack of a blazar within the {\it EGRET} localization
errors, like 3EG\,J0241+6103, 3EG\,J1824-1514 and
GRO\,J1834-04.

One of these has emerged as a new type of $\gamma$-ray source with the rediscovery of 3EG\,J0241+6103 (COS-B 2CG\,135+01), which was associated, although the position was uncertain, with LSI\,$+61 \adeg 303$, a "prominent radio flaring star system'', high mass X-ray binary (HMXB) system at 2 kpc constituted of a B0\,Ve star and a neutron star orbiting on an excentric orbit with a 26.5 days period. A daily/monthly variability was seen with {\it EGRET}, but no periodicity detected\cite{tavani:1998}. A TeV source has then been detected by MAGIC and then VERITAS, at this position, with a periodic signal modulated at the orbital period. {\it Fermi} has detected a $\gamma$-ray source, 0FGL\,J0240.36113, at the position RA=40.076, DEC=61.233 with a 95\% error radius of $1.8 \amin$, consistent with the optical counterpart, and with a periodicity of $26.6 \pm 0.5$\,days, the emission peaking at the periastron, and a spectrum reminiscent of the pulsars, therefore definitely identifying in the MeV-GeV domain the first $\gamma$-ray binary source with the HMXB system\cite{abdo:2009ApJ701L123}.

The second interesting source is the case of 3EG\,J1824-1514, detected by {\it EGRET}, but without modulation, as spatially coincident with LS\,5039, an HMXB constituted of a likely neutron star orbiting around an O6.5 star. HESS detected a periodic signal modulated at the orbital period of 3.91 days. {\it Fermi} firmly detected at more than $12 \sigma$ a periodic source, modulated at $3.91 \pm 0.05$ days, in a complicated region with a very intense Galactic diffuse emission.

The most dynamic example from {\it EGRET} is GRO\,J1838-04 (3EG\,J1837-0423), which produced an intense outburst in June 1995\cite{tavani:1997}. The flux above 100\,MeV in a 3.5 day period was found to be a factor of 7 brighter than in later observations of the region. Notably, no blazar counterpart is known within the 99\% {\it EGRET} error contour. The absence of a flat spectrum radio source at the levels typical of the {\it EGRET} blazars made this a candidate for a different type of $\gamma$-ray emitter. The proximity of such a unique outburst to the inner Galaxy led to speculation of a possible Galactic origin. The question of the progenitor of this activity remains as well as the broader question of the existence of similar sources of this type. As stated in Ref.~\refcite{tavani:1997}, ``other unidentified {\it EGRET} sources near the GP appear to be time variable with V$>1.5$'' in Ref.~\refcite{mclaughlin:1996ApJ473.763}.

Finally, {\it Fermi} has recently detected a variable high energy source coinciding with the position of the X-ray binary and microquasar Cygnus X-3, modulated at its short orbital period of 4.8 hours. Cygnus X-3 is an HMXB system located at a distance of $\sim 7$\,kpc, with a compact object of nature still matter in the debate, orbiting a Wolf-Rayet star\cite{abdo:2009Science326.1512}. 

\section{LAT Detection of $\gamma$-ray transients} \label{detection}

The {\it Fermi} Large Area Telescope (LAT) is very well suited
for monitoring variability in the GeV sky. The
large effective area ($>8000$\,cm$^2$ on axis above 1\,GeV),
wide field-of-view ($\sim 2.4$\,sr), and excellent angular resolution
(better than $1\adeg$ above 1\,GeV) greatly enhance
the sensitivity to transient activity in comparison to
previous $\gamma$-ray instruments in this energy range.
A notable departure from previous observations is that
the energy range of the LAT (20\,MeV to $>300$\,GeV)
extends above that covered by {\it EGRET}. The combination
of the wide field-of-view with the sky scanning observational
mode supplies coverage of the sky every $\sim 3$\,hours
(2 orbits). This enables the detection of fainter objects
in shorter intervals than previously possible. 
Also, the localizations for sources above threshold are much better than $1\adeg$ even on short timescales. These capabilities are critical for triggering rapid multiwavelength follow-up observations of LAT transients. The LAT has so far detected three transient events near the GP that have not been associated with blazars\footnote{There is a fourth transient event, Fermi\,J0109+6134, detected on 1st of February 2010\cite{vandenbroucke:2010}, which is probably a blazar shining through the Galactic plane; therefore we will not report more about this source in this paper focused on Galactic transients.}. Here, we report on the $\gamma$-ray characteristics and the multiwavelength follow-up observations that were triggered shortly after the detections.

All the unidentified transients were first detected
by the LAT automated science processing (ASP)\cite{chiang:2007}.
The ASP flare search locates and analyzes detected
point sources in the LAT photon data as these become
available for processing on the ground. The search runs
on 6-hour, 1-day, and 1-week intervals. 
The latency between the time the data are acquired and when they are
available for ground processing is short enough to allow
alerts to be communicated and follow-up multi-wavelength observations
triggered within a day. The LAT team continuously
monitors the output of the automated searches for new
source detections and flares from known LAT sources.
The automated processing reports candidate locations
and flux estimates, a summary is given in weekly reports in {\it http://fermisky.blogspot.com}.
The transients presented here are high-confidence
detections that appeared in multiple 6 hour and daily
ASP searches.

     \subsection{3EG\,J0903-3531}

\begin{figure}[ht]
\begin{center}
  \psfig{figure=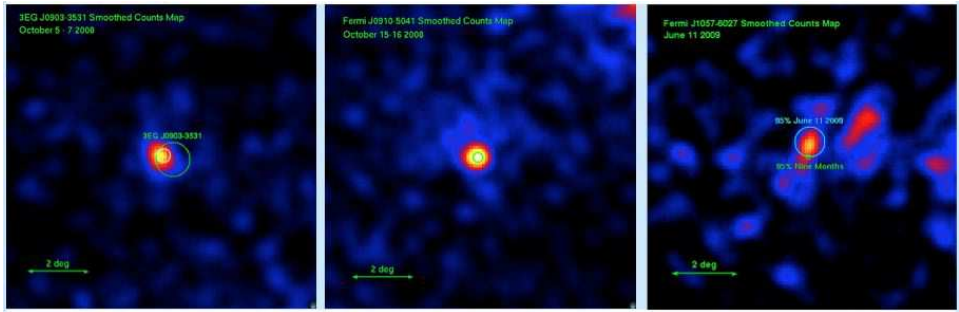,width=5in}
  \end{center}
  \caption{{\it Left:} LAT counts map (E$>100$\,MeV) in celestial coordinates of
3EG\,J0903-3531 from 5 October through 7 October 2008. The image
is smoothed with a $0\adegp5$ width Gaussian. The green circle shows
the 3EG 95\% error circle. The white circle represents the LAT 95\%
error circle based on the preliminary analysis of the flare period. 
{\it Middle:} LAT counts map (E$>100$\,MeV) in celestial coordinates of
J0910-5041 from 15 October through 16 October 2008. The image is
smoothed with a $0\adegp5$ width Gaussian. The green circle represents
the LAT 95\% error circle based on the preliminary analysis of the flare
period.
{\it Right:} LAT 95\% confidence region for the flare (cyan) and the nine-month source position (green) for Fermi\,J1057-6027. 
}
  \label{fig1}
\end{figure}

LAT observations on 6 October 2008 revealed a newly
detected LAT source spatially associated with 3EG\,J0903-3531\cite{hays:2008} in the 3EG catalog\cite{hartman:1999} (see Fig.~\ref{fig1} left panel). The preliminary
localization (J2000.0: RA$=136\adegp25$, DEC$=-35\adegp45$,
$r68=0\adegp12$; this is a correction to the originally reported ATel value) 
was based on one day of observations.
Additional analysis demonstrated the source was detected
(TS$>25$, $\sim 5 \sigma$) on 5, 6, and 7 October 2008
before falling below the daily sensitivity threshold for
this region. The LAT flux (E$>100$\,MeV) reported for
the flare exceeded $10^{-6}$\,photons\,cm$^{-2}$\,s$^{-1}$, which is about a
factor of 5 greater than the {\it EGRET} flux, and the source brightened by a factor of 15 in 3 days.
Despite being bright over several days, this source did
not exceed the $10 \sigma$ criteria for the LAT bright source list\cite{abdo:2009ApJS183.46}, which includes data from 4 August 2008 through 30
October 2008.

     \subsection{Fermi\,J0910-5041}

This source was first detected on 15 October 2008\cite{cheung:2008} (see Fig.~\ref{fig1} middle panel). It appears in the LAT bright source list as 0FGL\,J0910.2-5044\cite{abdo:2009ApJS183.46} and is flagged as variable. The preliminary
localization, (J2000.0: RA, Dec$=137\adegp69$,
$-50\adegp74$, r68$=0\adegp07$), is based on one day of data
and only accounts for statistical errors. The reported flux
above 100\,MeV exceeded $10^{-6}$\,photons\,cm$^{-2}$\,s$^{-1}$.
The bright source list analysis includes three months
of observations and uses a more detailed prescription
for calculating the 95\% error circle that also includes
systematic effects (see Ref.~\refcite{abdo:2009ApJS183.46}). The preliminary coordinates
are consistent with the bright source list position, providing
an additional confirmation of the LAT capability
for consistent localization on relatively short timescales.


\subsection{Fermi\,J1057-6027}

The LAT detected a new transient $\gamma$-ray source in the GP with exhibited an outburst on June 11, 2009, with a flux (E$>100$\,MeV) of $2.4 \pm 0.7 \times 10^{-6}$\,photon.s$^{-1}$cm$^{-2}$ (statistical)\cite{yasuda:2009} (see Fig.~\ref{fig1} right panel). 
The flare had the shortest duration and softest spectrum of the unassociated, low latitude transients making it less significant ($\sim 5 \sigma$) over foreground and background emission.
It coincided with a LAT source detected in 9-months (Aug 2008 - Apr 2009) of all-sky monitoring data (J2000.0: RA$ = 164\adegp308$, Dec =$-60\adegp458$, with a 95\% confidence error circle radius $0\adegp07$ (statistical)). The new LAT detection increased by a factor of $\sim 10$ with respect to the source flux level detected during the earlier period. There is no previously reported {\it EGRET} $\gamma$-ray detection at this location.

\section{Multiwavelength Observations} \label{multiwavelength}

Following each LAT transient detection, we triggered
target of opportunity {\it Swift}\cite{gehrels:2004} observations designed
to search for plausible counterparts. A pair of snapshot
(4-7\,ks) exposures were obtained approximately two to three days after the onset of $\gamma$-ray activity for the targets, with a third observation obtained 2 weeks
(3EG\,J0903-3531) and 1 month later (J0910-5041) in
order to search for variability on longer timescales.

\begin{figure}[ht]
\begin{center}
\psfig{file=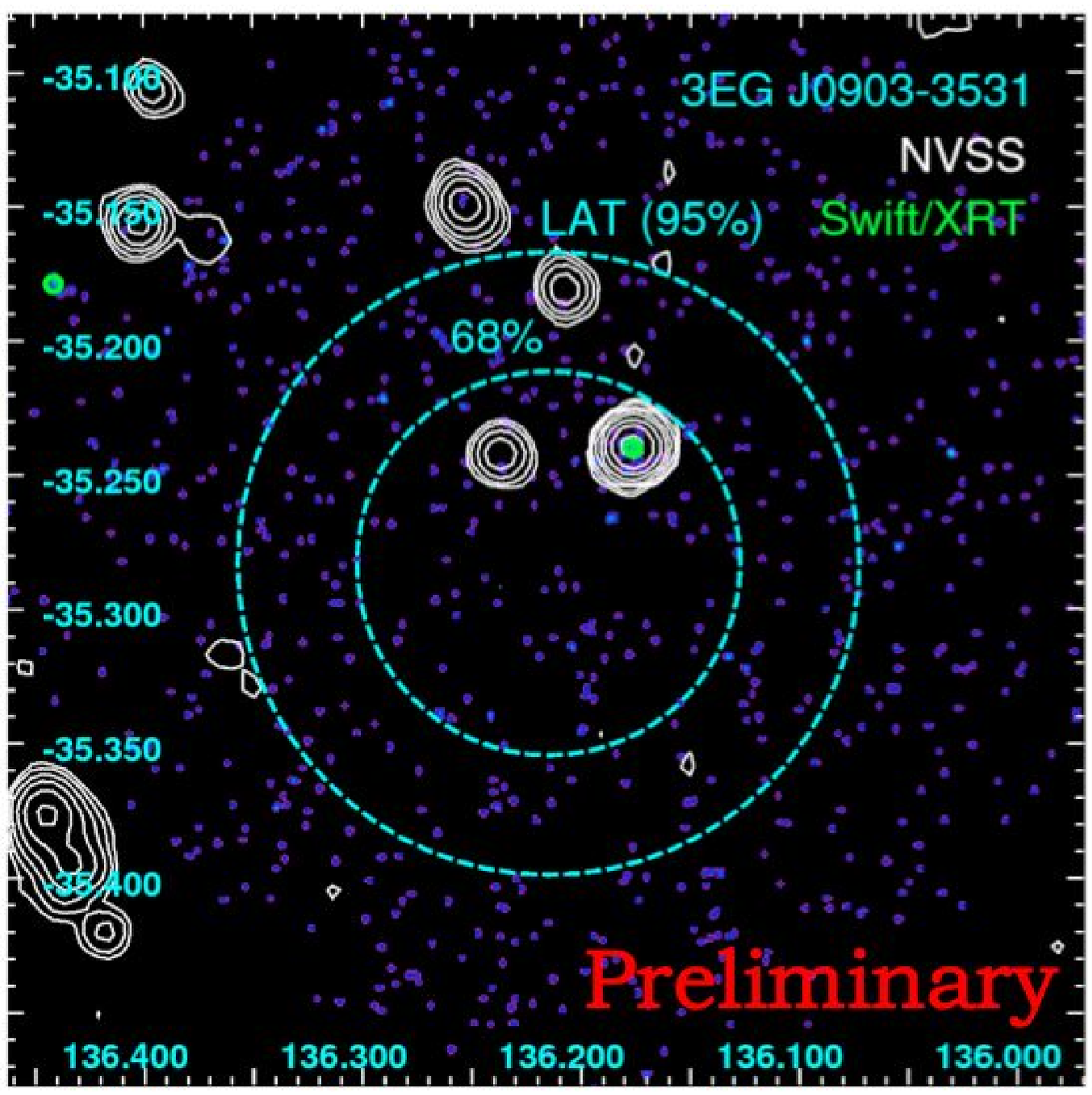,width=2.4in}
\psfig{file=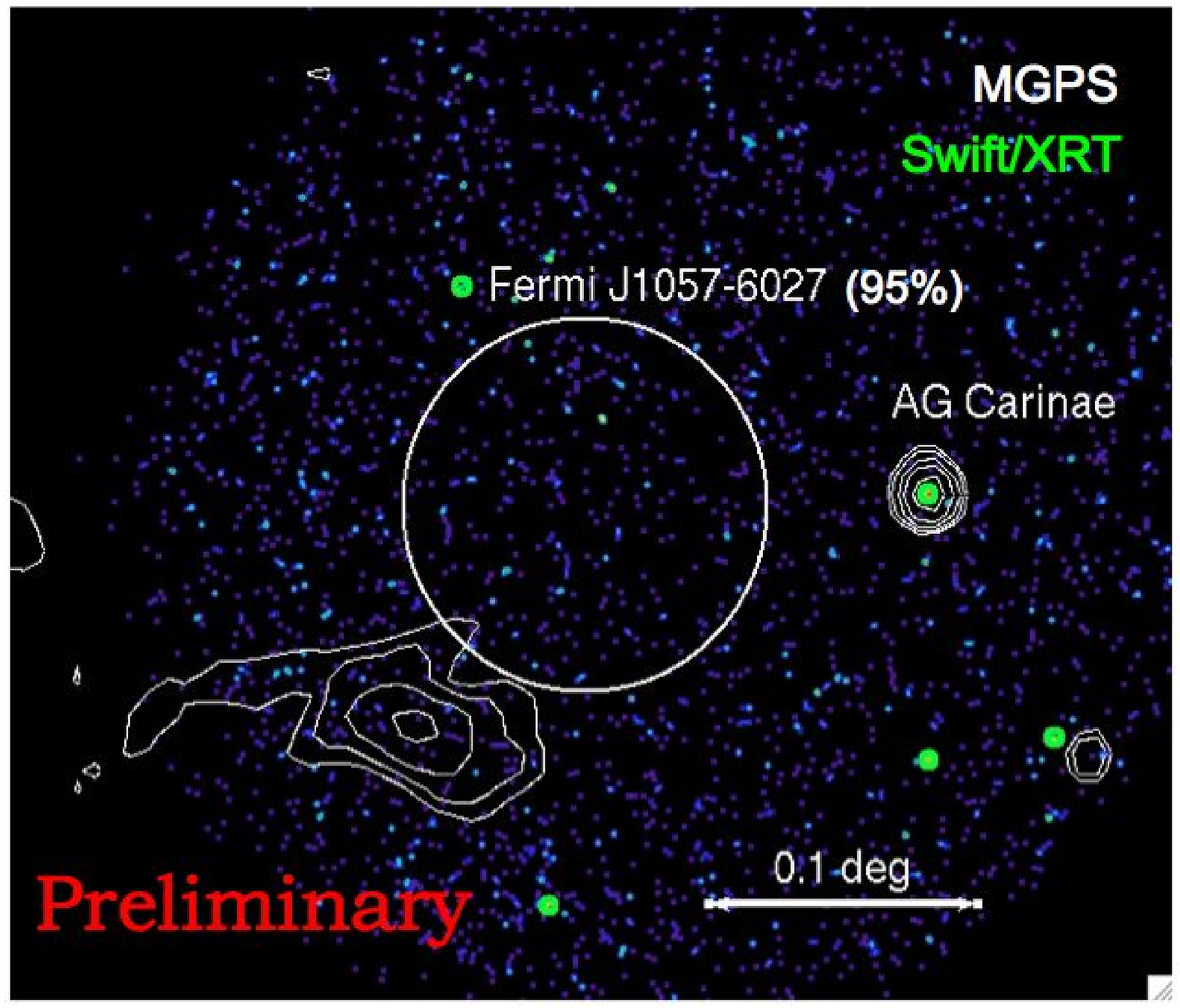,width=2.5in}
\end{center}
\caption{{\it Left:} Archival radio and {\it Swift}/XRT images of the field containing 3EG\,J0903-3531. {\it Swift}/XRT observations of the new LAT position (cyan) revealed a likely X-ray counterpart that coincides with a flat-spectrum radio object. {\it Right:} {\it Swift}/XRT image of the field containing Fermi\,J1057-6027. The white circle marks the 95\% error circle for the LAT detection.
}
\label{multi1}
\end{figure}

Within the LAT r68 of 3EG\,J0903-3531, there are
four $2.4-4.0 \sigma$ X-ray sources from the summed XRT
exposure (Fig.~\ref{multi1} left panel). The observed (unabsorbed) 0.3-10
keV fluxes are in the range $2.7 (5.3) - 6.3 (12.5) \times 10^{-14}$\,erg.s$^{-1}$cm$^{-2}$, assuming $\Gamma=2$ and Galactic absorption
$= 2.6 \times 10^{21}$\,cm$^{-2}$. There are no radio counterparts
to the X-ray sources at the depth of the 1.4\,GHz NVSS\cite{condon:1998} image, although two faint (2.5-3.9 mJy at 1.4 GHz)
radio sources are found within r68 (NVSS\,J090428.87-353007.5, NVSS\,J090458.76-353145.4)\cite{hays:2009}.

\begin{figure}[ht]
\begin{center}
\psfig{file=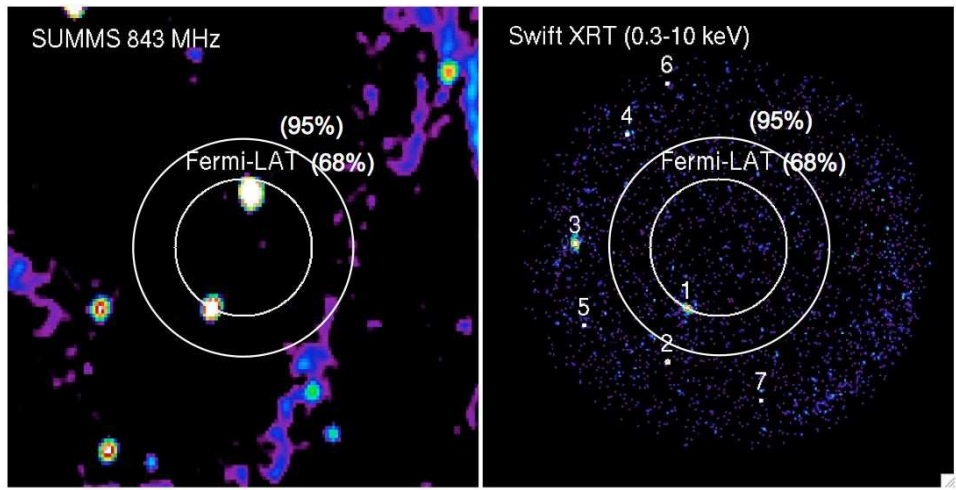,width=5in}
\end{center}
\caption{Archival radio (left panel) and {\it Swift}/XRT (right panel) images of the field containing Fermi\,J0910-5041. The two white circles mark the 68\% and 95\% error circles for the LAT detection.}
\label{multi2}
\end{figure}

In the case of J0910-5041, a single X-ray source
(Swift\,J091057.47-504808.5) was detected within the
LAT r68 (Fig.~\ref{multi2} right panel) . The observed (unabsorbed) 0.3-10
keV flux of $3.2 (13.3) \times 10^{-13}$\,erg.s$^{-1}$cm$^{-2}$, assuming $\Gamma=2$ and Galactic absorption = $1.3 \times 10^{22}$\,cm$^{-2}$. Ref~\refcite{landi:2008} found weak evidence for X-ray variability in
this source and that it has a radio counterpart detected
in the SUMMS image\cite{bock:1999} with a flat-spectrum\cite{sadler:2008}.
An additional radio source is found within the LAT r68
(SUMMS\,J091042-504103)\cite{hays:2009} (Fig.~\ref{multi2} left panel).

Finally, in the case of Fermi\,J1057-6027, a target of opportunity observation by {\it Swift}/XRT on June 13, 2009 did not detect any source within the $\sim 24\amin$ field-of-view of XRT\cite{torres:2009,cheung:2009}, and no radio counterparts have been found either (see Fig.~\ref{multi1} right panel).

\section{Conclusion}

The task of identifying counterparts for unidentified transients remains challenging. Although greatly reduced from {\it EGRET}, the LAT error circles remain large compared to the resolution of telescopes in other wavebands. Additional arguments based on temporal and spectral characteristics are required to support firm associations. Ultimately, identifications of the Galactic transients require observations of related variability between the $\gamma$-ray source and a candidate Galactic counterpart at lower frequency. In the absence of a detection of significant activity of the potential radio and X-ray counterparts for the LAT transients, they remain unidentified. These sources continue to be monitored regularly for $\gamma$-ray activity as a part of the {\it Fermi} sky survey observations.

\section{Acknowledgements}

The {\it Fermi}-LAT Collaboration acknowledges support
from a number of agencies and institutes for both the
development and operation of the LAT as well as the scientific data analysis. These include NASA and the DOE
in the United States, the CEA/Irfu and IN2P3/CNRS
in France, ASI and INFN in Italy, MEXT, KEK, and
JAXA in Japan, and the K.A. Wallenberg Foundation,
the Swedish Research Council and the National Space
Board in Sweden. Additional support from INAF in Italy
for science analysis during the operations phase is also
gratefully acknowledged.
This work was supported by the Centre National d'Etudes Spatiales (CNES), based on observations obtained through MINE: the Multi-wavelength INTEGRAL \& Fermi NEtwork.

\bibliographystyle{ws-procs975x65}

\end{document}